\documentclass[11 pt,final,journal,letterpaper,oneside,onecolumn]{IEEEtran}
\usepackage{physics}
\usepackage{amsmath}
\usepackage{amsfonts}
\usepackage{mathtools}
\usepackage{amsfonts}
\usepackage{amssymb}
\usepackage{array}
\usepackage{makeidx}
\usepackage{csquotes}
\usepackage{graphicx}
\usepackage{cite}
\usepackage{xcolor}
\usepackage{subcaption}
\usepackage{graphicx}
\usepackage{multicol}
\usepackage{epstopdf}
\usepackage{multirow}
\usepackage{pifont}
\begin{document}
%
\title{Opportunities for Intelligent Reflecting Surfaces\\ in 6G-Empowered V2X Communications}

\author{Wali Ullah Khan, \textit{Member, IEEE}, Asad Mahmood, Arash Bozorgchenani, \textit{Member, IEEE}, \\Muhammad Ali Jamshed, \textit{Senior Member, IEEE}, Ali Ranjha, Eva Lagunas, \textit{Senior Member, IEEE},\\ Haris Pervaiz, \textit{Member, IEEE}, Symeon Chatzinotas, \textit{Senior Member, IEEE},\\ Bj\"orn Ottersten, \textit{Fellow, IEEE}, and Petar Popovski, \textit{Fellow, IEEE} \thanks{Wali Ullah Khan, Asad Mahmood, Eva Lagunas, Symeon Chatzinotas, and Bj\"orn Ottersten are with the Interdisciplinary Center for Security, Reliability and Trust (SnT), University of Luxembourg, 1855 Luxembourg City, Luxembourg (emails: \{waliullah.khan,asad.mahmood,eva.lagunas,Symeon.Chatzinotas,bjorn.ottersten\}@uni.lu).

Arash Bozorgchenani and Haris Pervaiz are with the School of Computing and Communications, Lancaster University, United Kingdom, (email: a.bozorgchenani@lancaster.ac.uk, h.b.pervaiz@lancaster.ac.uk).

Muhammad Ali Jamshed is with the with James Watt School of Engineering, University of Glasgow, Glasgow, G12 8QQ, United Kingdom (e-mail: muhammadali.jamshed@glasgow.ac.uk)

A. Ranjha is with the Department of Electrical Engineering, \'Ecole de Technologie Sup\'erieure, Montr\'eal, QC, H3C 1K3, Canada, (e-mail: ali-nawaz.ranjha.1@ens.etsmtl.ca).

Petar Popovski is with the Department of Electronic Systems, Aalborg University, Denmark (email: petarp@es.aau.dk).

}}%

\maketitle

\begin{abstract}
The applications of upcoming sixth-generation (6G)-empowered vehicle-to-everything (V2X) communications depend heavily on large-scale data exchange with high throughput and ultra-low latency to ensure system reliability and passenger safety. However, in urban and suburban areas, signals can be easily blocked by various objects. Moreover, the propagation of signals with ultra-high frequencies such as millimeter waves and terahertz communication is severely affected by obstacles. To address these issues, the Intelligent Reflecting Surface (IRS), which consists of nearly passive elements, has gained popularity because of its ability to intelligently reconfigure signal propagation in an energy-efficient manner. Due to the promise of ease of deployment and low cost, IRS has been widely acknowledged as a key technology for both terrestrial and non-terrestrial networks to improve signal strength, physical layer security, positioning accuracy, and reduce latency. This paper first describes the introduction of 6G-empowered V2X communications and IRS technology. Then it discusses different use case scenarios of IRS enabled V2X communications and reports recent advances in the existing literature. Next, we focus our attention on the scenario of vehicular edge computing involving IRS enabled drone communications in order to reduce vehicle computational time via optimal computational and communication resource allocation. At the end, this paper highlights current challenges and discusses future perspectives of IRS enabled V2X communications in order to improve current work and spark new ideas.
\end{abstract}  


\IEEEpeerreviewmaketitle

\section{Introduction}

The sixth-generation (6G)-empowered vehicle to everything (V2X) communications is essential to smart city transportation systems. Robust wireless connections and cutting-edge sensors will completely transform the safety and comfort of the existing transportation systems \cite{gyawali2020challenges}. The future transportation industry will incorporate a wide range of technologies, including those for passenger and driver protection, autonomous driving, traffic management, and passenger amusement. By providing pervasive connectivity, secure data sharing, energy-efficient transmissions, and quick computation, 6G wireless technology is the backbone of transportation industry. In contrast to the 5G, which is all about autonomous driving, the 6G standard will be propelled by the need to ensure the safety of autonomous vehicles, facilitate the sharing of more comprehensive road traffic data, implement traffic planning using augmented reality (AR) and virtual reality (VR) technology, and support more sophisticated digital content and gaming applications. Furthermore, the 6G transportation system will offer terabit-per-second data rates, which are exceptionally high. As a result, the latency of wireless communications can be reduced to under 1 millisecond, and the packet delivery ratio can be increased to $\approx100\%$ \cite{khan2022noma}. 6G will be enabled by technologies including intelligent reconfigurable surfaces (IRS), terahertz communications (THz), blockchain, ambient backscatter communications, and artificial intelligence.

Besides the promise of the above features, V2X communications also faces several challenges. For example, shadowing effects can significantly impact the efficiency and effectiveness of V2X communications due to obstacles like buildings in urban settings or hills and trees in rural areas. Therefore, limited energy reservoirs and spectrum resources would be the main challenges for large-scale V2X communications in 6G. Future V2X communications may also suffer with low transmission latency, unreliable wireless connectivity, and/or limited coverage. Moreover, high velocity vehicles impacts channel stability, having a negative impact on data rates. Accordingly, changing the position of drones in the air complicates communication even further. Keeping a high degree of energy efficiency in V2X communications while attempting to control the propagation and fading of THz signals is an open question. Driving safety and communication security are compromised by V2X communications that are unstable. It is essential to increase the range of communication and strengthen it in a sustainable manner.

The IRS has been seen as a potentially game-changing technology in 6G, with the ability to manipulate signal propagation and develop an intelligent radio environment \cite{9424177}. Using reflection and programming, IRSs can alter the phase of incoming electromagnetic (EM) waves, allowing for the redesign of channels. IRS reflection can create a new propagation path around an obstacle that is impeding the direct Line-of-Sight (LoS) link between the source and destination. In conventional communication systems, re-engineering the transceiver is the only option for boosting system performance. The IRS adds a new design parameter to wireless networks. Therefore, IRS technology can be used to enhance vehicular communications and offer indirect LoS links that are both cost-effective and energy-efficient. Because significant performance gains are achieved only when the transceiver is close to the IRS, a permanently deployed IRS will limit its potential. Given the transient nature of vehicles, mobile IRS is viewed as a viable option for V2X networks.

Great potential exists for 6G V2X-empowered communications thanks to IRS's ability to enable beyond LoS and energy-efficient communications. The IRS promises to help vehicle to infrastructure (V2I), vehicle to vehicle (V2V), vehicle to drone (V2D), and vehicle to satellite (V2S) communications in 6G networks by improving multipath propagation and expanding transmission coverage in high frequency bands, i.e., millimeter wave (mmWave) and THz \cite{CaoVT22}. The IRS is also simple to deploy due to its two-dimensional plane surface structure. Furthermore, the passive reflection mechanism enables IRS to operate in a low-energy-consumption mode, meeting the green 6G-empowered V2X communications requirements. Recent hardware and material research indicate that it can control the reflection dynamically, allowing the IRS to perform real-time beamforming and serve multiple vehicles. IRS's ability to use reconfigurable passive beamforming to strengthen physical layer security in vehicular communications on the ground and in the air is a major advantage.

This paper describes the IRS opportunities in 6G-empowered V2X communications and highlights some existing problems for ground and aerial/space V2X communications. First, we discuss various use case scenarios of IRS enabled V2X communications and provide recent advances in the literature. Then, we present a case study on IRS-enabled cooperative drone communications in vehicular edge computing (VEC), with the goal of reducing vehicle computational time using a new optimization framework. Before IRS can be widely used in 6G-empowered V2X communications, some issues need to be resolved. We underline the difficulties so as to provide direction for the implementation of IRS in terrestrial and non-terrestrial V2X communications. The rest of this article is structured as follows: Section II contains use case scenarios as well as recent advances. Section III introduces a new optimization framework for minimizing computational time in the IRS-enabled VEC network. Section IV discusses unresolved issues and potential future research directions. Section V concludes with closing remarks.

\begin{figure*}[!t]
\centering
\includegraphics[width=0.90\textwidth]{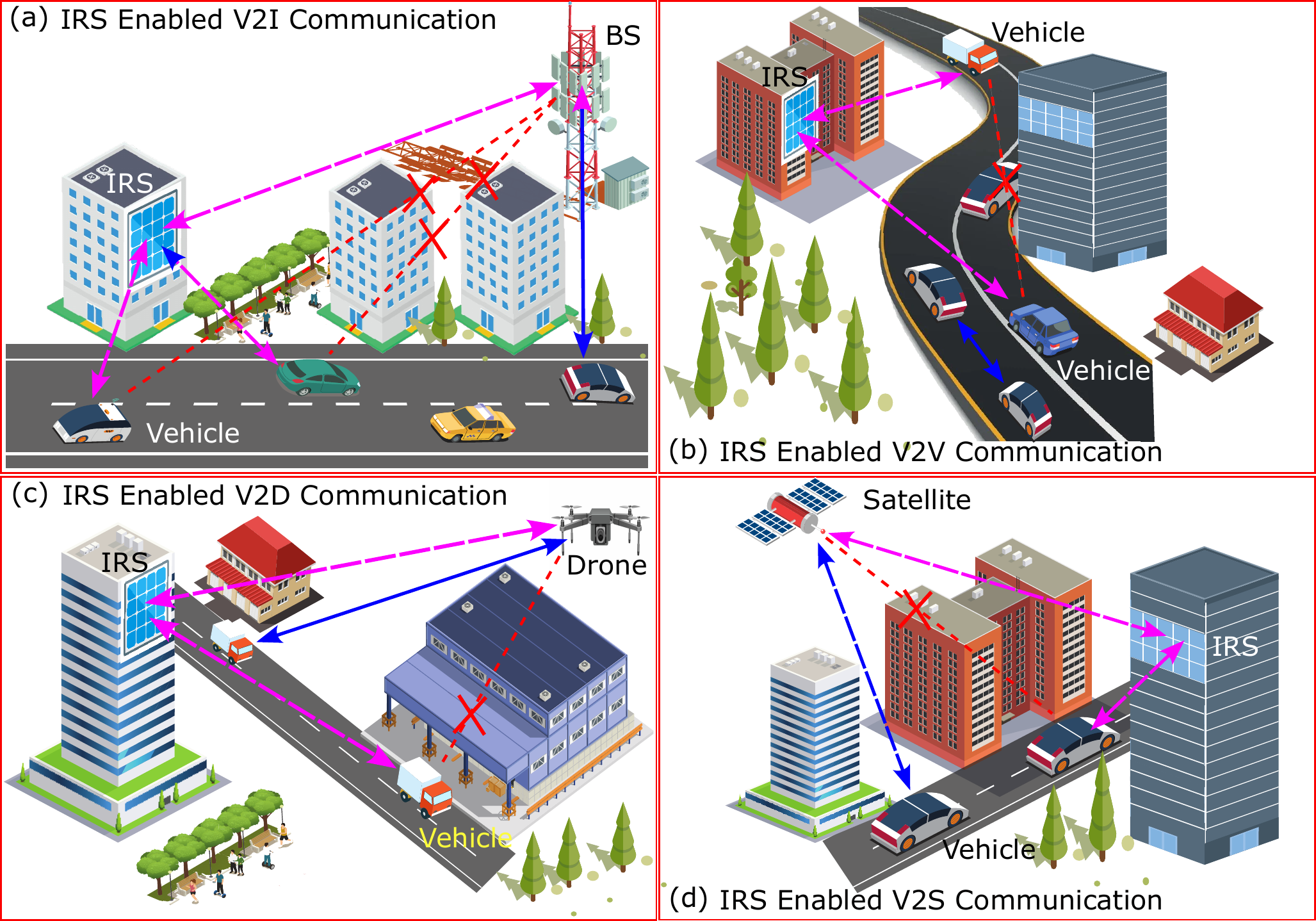}
\caption{Potential use case scenarios of IRS enabled V2X communications; (a) IRS enabled V2I communication, (b) IRS enabled V2V communication, (c) IRS enabled V2D communications, and (d) IRS enabled V2S communication.}
\label{blocky}
\end{figure*}
\section{IRS Enhanced V2X Communications: Use Case Scenarios and Recent Advances}
In this section, we first highlight and discuss different potential use case scenarios in next-generation IRS enabled V2X communications. Then we study and report recent advances in IRS enabled V2X communications existed in the literature. In addition, we also present the comparison table of these works.

\subsection{Use Case Scenarios of IRS enabled V2X Communications}
The integration of IRS into V2X communications can significantly improve the system performance. These improvements can be achieved in form of security, capacity, energy efficiency and coverage extension. Some of the potential use case scenarios of V2X communications involving IRS are shown in Fig. 2. Specifically, these use cases are V2I, V2V, V2D and V2S, respectively. Moreover, these use case scenarios are discussed in the following subsections.  

\subsubsection{IRS Enabled V2I Communications}
V2I communications can face the challenges of signal blockage and large-scale fading in urban areas. One of the traditional methods in such use case scenario is to deploy relay devices to improve the received signal strength. However, it requires extra power consumption. The ability of the IRS to intelligently reconfigure the signal toward the receiver can extend the wireless coverage and enhance the system performance in non-line-of-site communications scenarios without consuming any energy. Figure \ref{blocky} (a) shows a V2I communications network where a base station (BS) communicates with multiple vehicles in an urban area. We can see that some vehicles face signal blockage due to high buildings, affecting the system's performance. In such a scenario, vehicles with non-line-of-site (nLOS) communications can be efficiently assisted by IRS to enhance the received signal strength and system performance. IRS can be efficiently mounted on a strategic position such as a high building to deliver the signal from BS to vehicles. Moreover, the vehicles with LOS communications can also benefit from IRS and receive their signal through IRS to further enhance their capacity. 

\subsubsection{IRS Enabled V2V Communications}
In the V2V use case scenario, the communications between different vehicles can be affected by other vehicles on the road and objects on the roadside. Specifically, the signal transmitted from one vehicle to another vehicle can face blockage and large-scale fading due to other vehicles and objects on the road, affecting the communications quality of services between two vehicles. Figure \ref{blocky} (b) provides a V2V communications network where multiple vehicles communicate with each other. The proposed scenario shows that the transmissions between two vehicles are blocked by a high building and vehicle, weakening their channel conditions. IRS can play a crucial role in assisting the signal delivery between two vehicles and enhancing their quality of service communications. IRS can be efficiently installed on the building wall to provide energy-efficient and secure reflection for incident signals towards the desired vehicle. Other vehicles can receive information signals through direct and RIS enabled communications links. 

\subsubsection{IRS Enabled V2D Communications}
The mobility of Drone can be efficiently used for communications in densely crowded environments such as large cities, sports grounds and other public gatherings to improve the connectivity, system performance and reduce the terrestrial network overhead \cite{khan2022opportunities}. In particular, the large-scale vehicles and other moving objects on roads in big cities can face several issues of performance, connectivity, fading and transmission latency. It puts an extra burden on the communications network due to a large-scale exchanged data among different vehicle and other objects on the roadside. In such a communication scenario, drones can be operated with the collaboration with IRS to improve the performance of vehicles and other objects on the road as shown in Figure \ref{blocky} (c). Specifically, IRS can assist the signal of those vehicles and object which faces nLOS connectivity/ signal blockage, large-scale fading, low data rates and high transmission latency. In the provided figure, the drone acts as a BS and IRS is mounted on the building to assist the signal. Moreover, drone can also be equipped with IRS to efficiently deliver the signal of a terrestrial network in high building areas with signal blockage.

\subsubsection{IRS Enabled V2S Communications}
Due to the mega low orbit constellation, satellite communications have recently gained significant attention
for supporting a wide range of services throughout the globe. Moreover, satellite communications intend to provide massive connectivity, high-speed data rate, and low transmission latency. It can be achieved by deploying large-scale, low-cost small satellites in low orbit. Furthermore, these satellites will play a key in the successful deployment of future autonomous vehicle networks. However, the high mobility of vehicles, the effect of obstacles in the urban areas, and shadowing between vehicles and satellites can disrupt the LOS connection and significantly reduce system performance. IRS can be efficiently deployed in these situations to deliver signals between vehicles and satellites successfully. As highlighted in Figure \ref{blocky} (d), two vehicles on the road are shown to communicate with the satellite. One vehicle is accessing signal directly from satellite through the LOS link while the other is facing the nLOS link and signal blockage due to obstacle. To address this issue, an IRS is strategically mounted on the building to assist the satellite signal. In addition, IRS can also be used in terrain and mountain areas for efficient satellite communications.  

\begin{table*}[tbp]
\centering
\caption{Recent advances in IRS enhanced V2X communications}
\label{Rel_Works}
\scriptsize
\begin{tabular}{ m{0.5cm}  m{5cm} m{1.5cm}  m{1.5cm}  m{3.2cm} m{3.7cm} } 
  \hline
  	\hline
  \textbf{Ref.} & \textbf{Use case scenario} & \textbf{IRS position} & \textbf{Transmission} &\textbf{Proposed solution method} & \textbf{Performance gain (objective)}\\
  \hline
  \cite{LiU_JSAC21}& Drones equipped with multiple antennas communicate with single antenna ground mobile users through IRS & On building & V2I & Decaying DQN & Minimizing energy consumption\\
   \hline
 \cite{Zhu_TETC22} & Task offloading of Vehicles to RSU through IRS, where RSU equipped with edge computing & On building & V2I & Dynamic task scheduling algorithm & Maximizing average offloading rate, successfully
computing rate, and successfully finish rate\\
    \hline
\cite{YuanbinTVT20} & BS communicates with vehicles through IRS and inter-vehicle communication & On building & V2I \& V2V & Alternating optimization & Maximizing sum capacity\\
	\hline
\cite{Pan_AI22}& Multiple vehicle clusters, each cluster consists of one head vehicle and multiple member vehicles. Processing requests from head vehicles to BSs through IRS & On building & V2I \& V2V & DRL algorithm & Optimizing energy efficiency and latency\\
   \hline
\cite{WangGC20}& RSU communicates with vehicles through IRS & On building & V2I & Series expansion and central limit theorem & Reducing outage probability\\
   \hline
\cite{AlHilo_TVT22}& RSU communicates with vehicles lying in dark zone through IRS & On building & V2I & DRL \& BCD algorithms & Maximizing the minimum average bit-rate\\
   \hline
\cite{Chen_TWC22}& Vehicles communicate with BS through RIS, where BS is equipped with multiple antennas & On building & V2I & Alternating optimization & Maximizing average sum-rate \\
     \hline
\cite{Chen_TWC21}& Vehicles communicate with BS through RIS and vehicles directly communicate directly with each other & On building & V2V \& V2I & BCD algorithm & Maximizing sum capacity \\
   \hline
\cite{Haung_ICC21}& BS communicates with vehicle through IRS, where IRS is mounted on vehicle & On vehicle & V2I & Heuristic transmission protocol \& Passive beamforming & Maximizing achievable rate\\
   \hline
\cite{Ai_TVT21}& (i) Vehicle communicates with vehicle through IRS in the presence of a passive eavesdropper, and (ii) IRS communicates with vehicle in the presence of a passive eavesdropper & On building & V2I \& V2V & Closed-form expression for secrecy outage probability & Improving secrecy\\
     \hline
[Our]& Vehicles communicate with access point of vehicular edge computing through IRS enable cooperative drone communication & On drone & V2I & SCA algorithm & Reducing computational time of vehicle task\\
     \hline
\end{tabular}
\end{table*}

\subsection{Recent Advances in IRS Enabled V2X Communications}
Although the application of IRS in wireless communication systems has been widely addressed, its application in vehicular environments has been briefly studied in the literature. In the following, we concisely review the most related works. 

For example, the work in \cite{LiU_JSAC21} has proposed a paradigm for exploiting IRS with non-orthogonal multiple access (NOMA) in drone-enabled wireless networks. The authors simultaneously optimize the phase shift control, dynamic trajectory design, signal decoding order, and power control to minimize the drone's energy consumption. They design a decaying deep Q-network (DQN)-based algorithm to solve the formulated problem. 
Simulation findings show that with RIS, the drone energy consumption may be significantly minimized, and IRS with NOMA consumes less energy than IRS with orthogonal multiple access.

Different from other works that mainly focused on communication, in \cite{Zhu_TETC22}, the authors have considered both communication and computation in a mobile edge computing enabled vehicular network involving IRS. They address the problem of task scheduling which includes allocating processor and link resources. They propose a dynamic task scheduling algorithm to maximize the computation throughput. The simulation results demonstrate the effectiveness of their approach in terms of task offloading rate, computing rate, and successful finish rate. Resource allocation for IRS enabled V2X communications has also been addressed in \cite{YuanbinTVT20}, where the authors aim to maximize the V2I capacity while guaranteeing minimum capacity of V2V links. They address a joint power allocation, IRS reflection coefficients, and spectrum allocation problem, proposing an alternating optimization algorithm.

The authors in \cite{Pan_AI22} design a similar 3-plane framework including vehicles, IRS-deployed buildings and BSs, which
perform the resource allocation and task processing for the vehicles. The vehicles' data is transmitted to the BSs through IRS. 
To reduce energy consumption and latency, the authors propose a deep reinforcement learning (DRL) strategy for efficient resource allocation, which includes the vehicle transmission power, IRS reflection phase shift, and BS detection matrix.

A downlink vehicular communication model is studied in \cite{WangGC20}. The authors consider communication from road side unit (RSU) to vehicles through IRS-enabled buildings when the vehicle becomes blocked by obstacle vehicles. They approximate the outage probability based on series expansion and central limit theorem. Finally, in the simulation results, they demonstrate improved vehicle coverage by using IRS and conclude that a larger IRS can reduce the outage probability.
In \cite{AlHilo_TVT22}, the authors consider an indirect transmission from RSUs through IRS deployed on buildings to some dark zones. Similar to \cite{WangGC20}, they utilize IRS as an indirect means of transmission due to the blockage of the vehicles. They formulate a joint resource scheduling and RSU passive beamforming aiming at maximizing the minimum average bit rate. 
They resort to DRL to obtain RSU wireless scheduling and Block Coordinate Descent (BCD) to solve the passive beamforming of the IRS.

In \cite{Chen_TWC22}, IRS has been integrated into millimeter wave (mmWave) vehicular communication. The authors address the issue risen by high mobility and the challenge of obtaining the accurate channel state information (CSI). 
They have studied both single and multi-vehicle cases with the objective of maximizing the uplink average achievable sum-rate by jointly optimizing the transmit power, multi-user detection matrix and the RIS reflection phase shift. In the numerical results, it is demonstrated that their transmission framework is robust to the performance loss caused by outdated CSI and keeps the average sum-rate at a favourable level. A similar problem is studied in \cite{Chen_TWC21}, where apart from the optimization parameters in \cite{Chen_TWC22}, they also consider the spectrum reuse of V2V links in the joint formulated mixed-integer non-convex optimization problem. Then they approximate the outage probability of the V2V links and 
decompose the problem into three sub-problems where the alternated updated problem achieves a near-optimal solution.
\begin{figure*}[!t]
\centering
\includegraphics[width=0.60\textwidth]{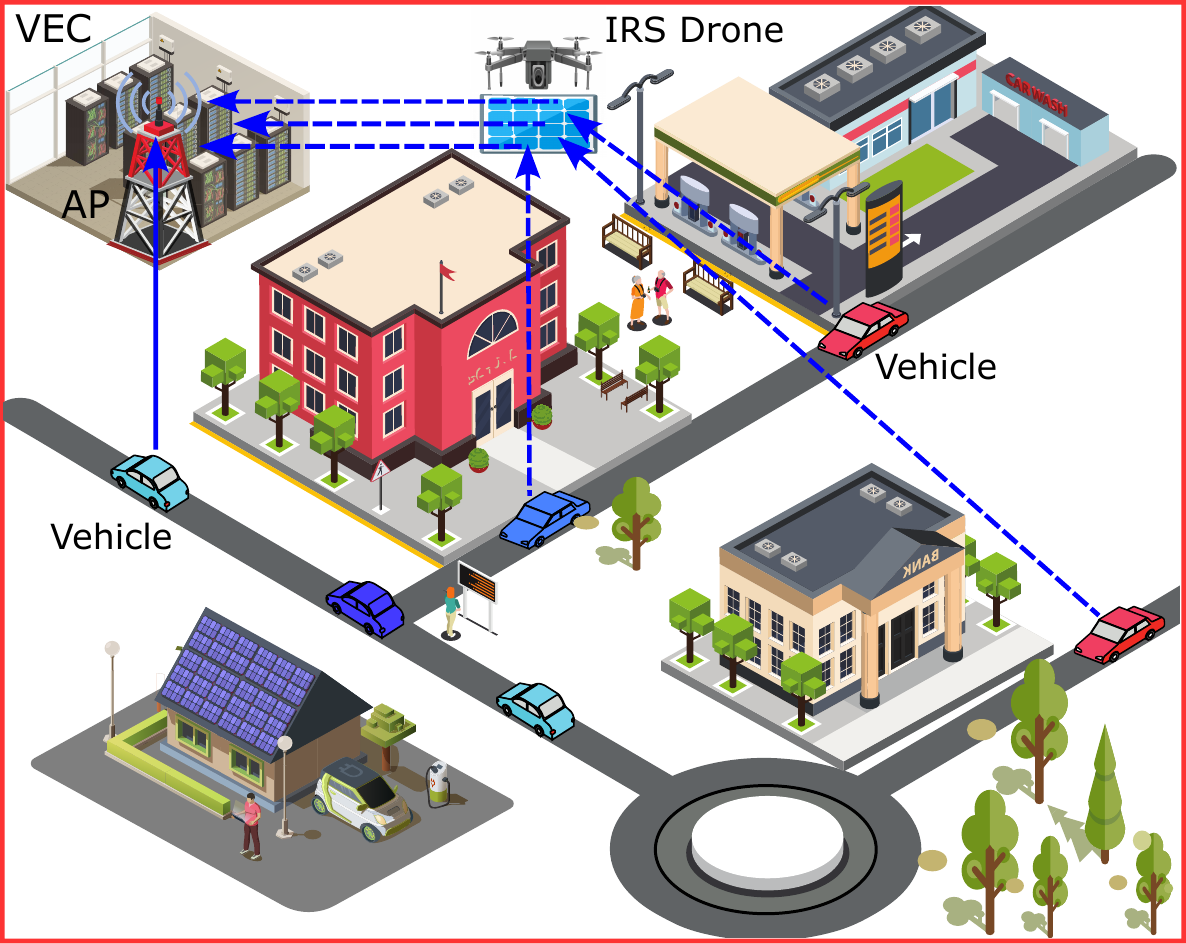}
\caption{System of vehicular edge computing empowered by IRS enabled drone communications}
\label{SM}
\end{figure*}
In \cite{Haung_ICC21} the authors study a high mobility communication scenario between passengers and BSs through IRS deployed on the vehicle. They address mitigating the Doppler effect through approximations and tuning the IRS reflection over time.


In \cite{Ai_TVT21}, the secrecy outage performance of RIS enabled vehicular communications is analyzed, where they consider two scenarios of V2V and V2I. In the V2V scenario the RIS functions as a relay and in the V2I case, it acts as a receiver. In both cases, the authors assume the presence of a passive eavesdropper. They derive the closed-form expression for the secrecy outage probability and demonstrate the effectiveness of the assistance of the RIS in both scenarios in the simulation results.

 Table \ref{Rel_Works} summarizes the most related works in the area of IRS enhanced V2X communications. Most of the works assume that IRS is intelligently deployed and configured under the control of a ground control station such as a BS. In most of the works, the IRS is considered to be deployed on the building sides, aiding the V2V and V2I vehicular communications. The above studies proved that the application of the IRS can boost the performance of vehicular communication by providing different example scenarios. 
Nevertheless, the research on IRS-aided V2X is still in its infancy as V2D, and V2S scenarios are not well-investigated and their capacities are not well exploited. In particular, in high mobility scenarios when IRS is equipped with drones or vehicles, resource allocation problems are even more challenging and worth investigating.

\section{Vehicular Edge Computing Empowered by IRS Enabled Cooperative Drone Communications} 
This section provides a new optimization framework for computational task minimization in vehicular edge computing (VEC) involving IRS-enabled cooperative drone communications. In the following, we first explain the system model, problem formulation, and the proposed optimization solution. Then we validate our proposed solution by presenting numerical results and their discussion.

\begin{figure*}[h!]
\centering
\begin{subfigure}{0.32\textwidth}
    \includegraphics[width=\textwidth]{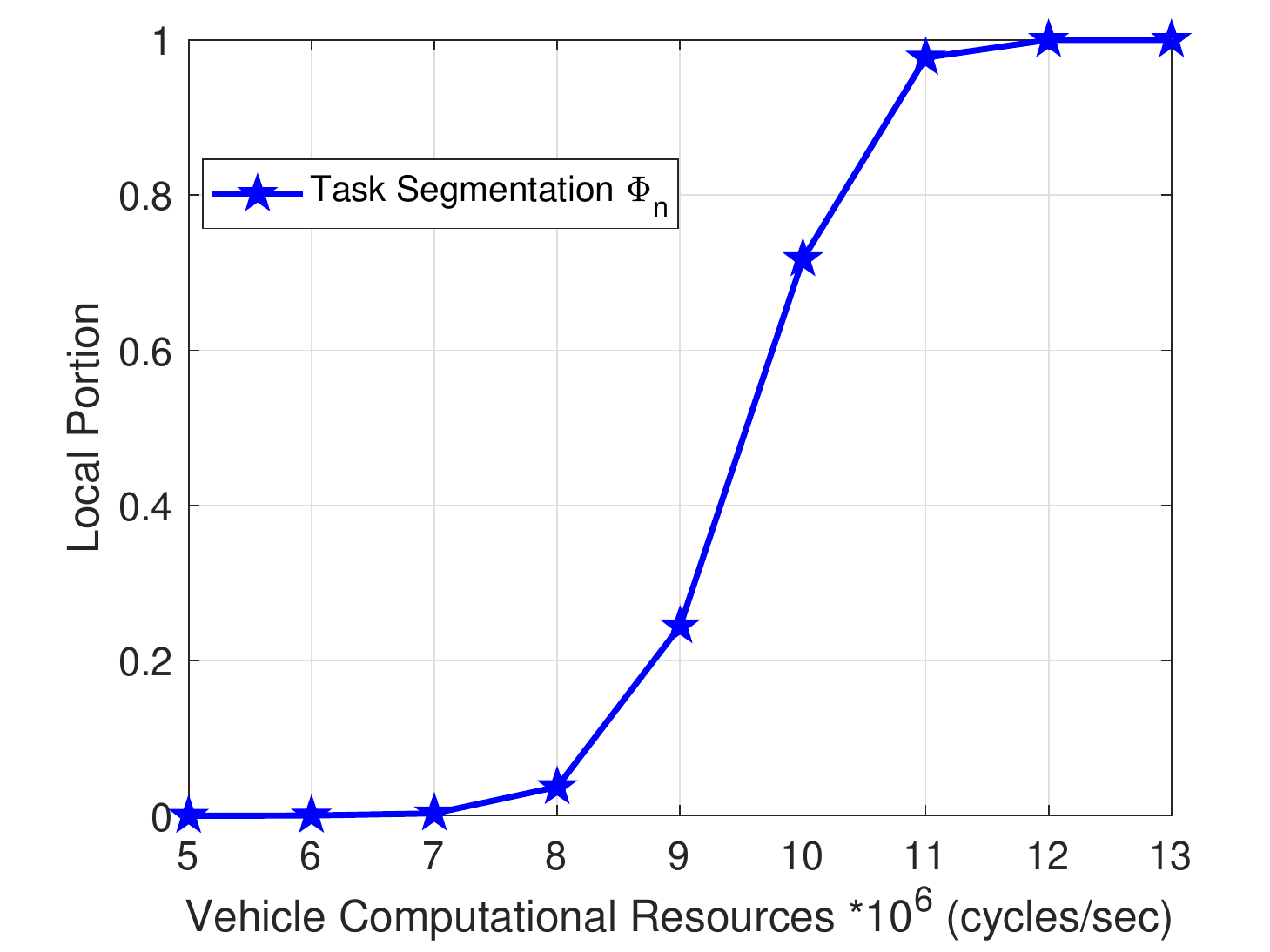}
    \caption{}
    \label{fig:first}
\end{subfigure}
\begin{subfigure}{0.32\textwidth}
    \includegraphics[width=\textwidth]{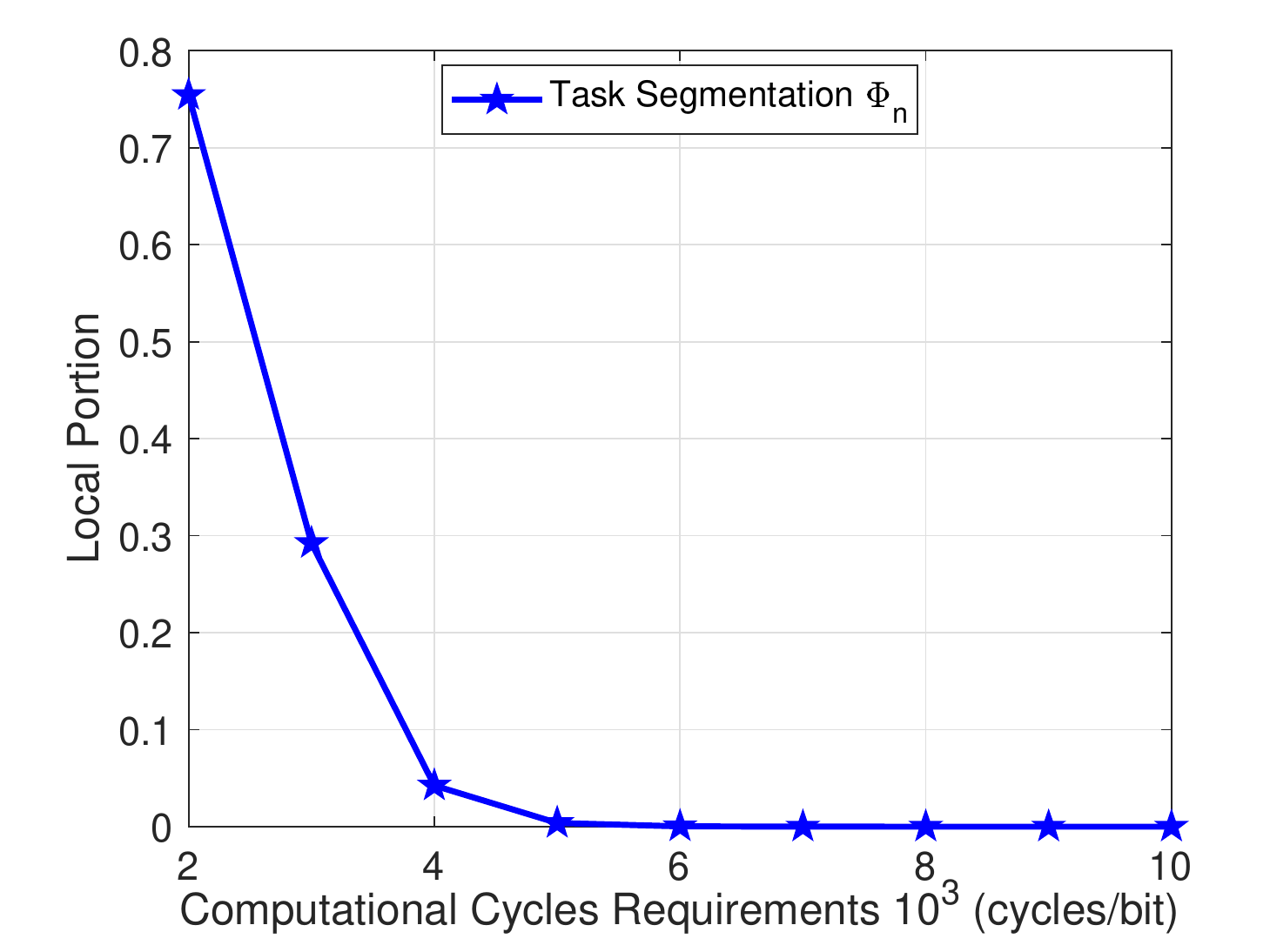}
    \caption{}
    \label{fig:second}
\end{subfigure}
\begin{subfigure}{0.32\textwidth}
    \includegraphics[width=\textwidth]{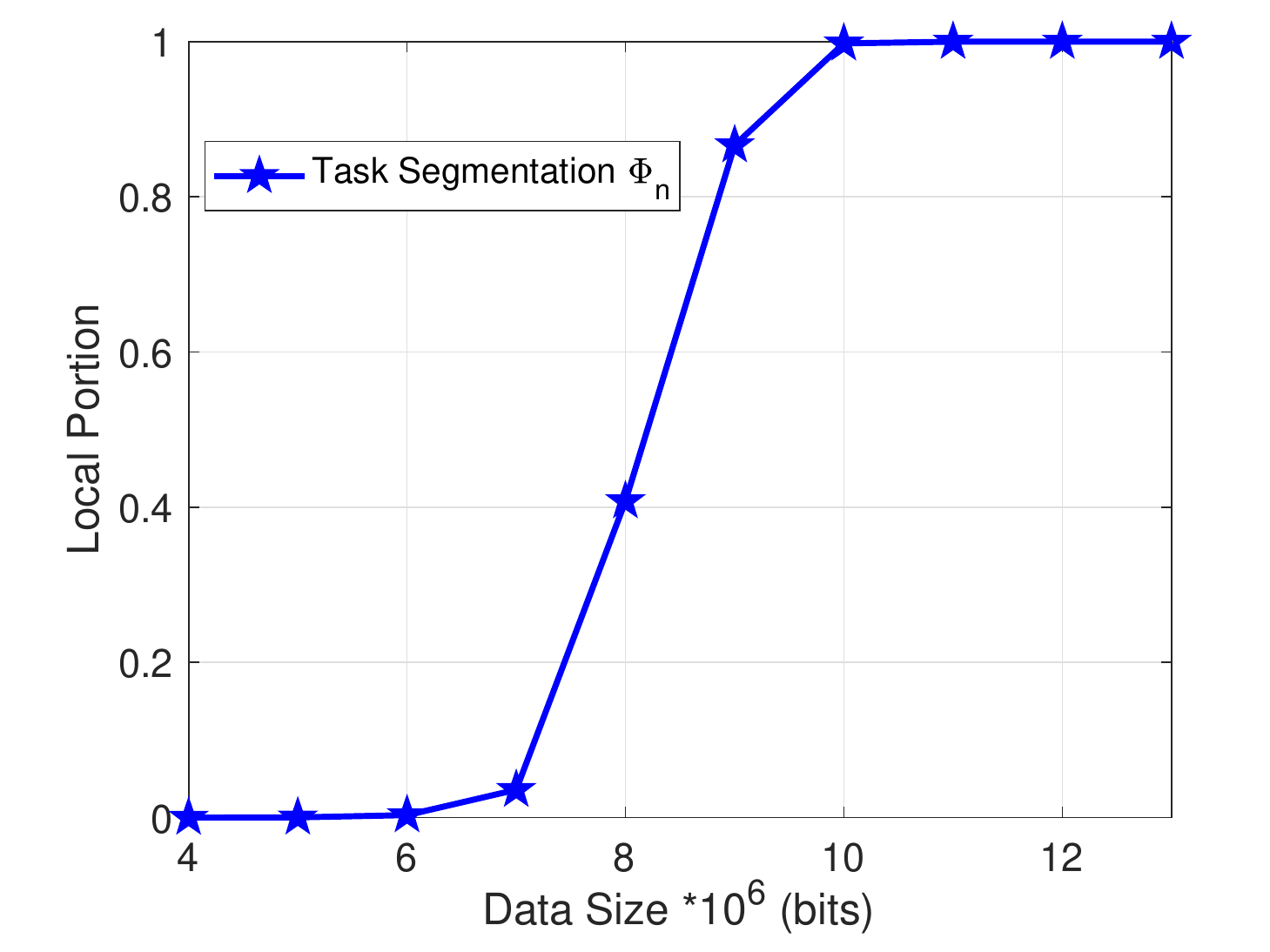}
    \caption{}
    \label{fig:third}
\end{subfigure}
\caption{Impact of Vehicle computational resources, Cycles requirement and data size of task offloading}
\label{fig:figures}
\end{figure*}

\begin{figure*}[!t]
\centering
\includegraphics[width=0.70\textwidth]{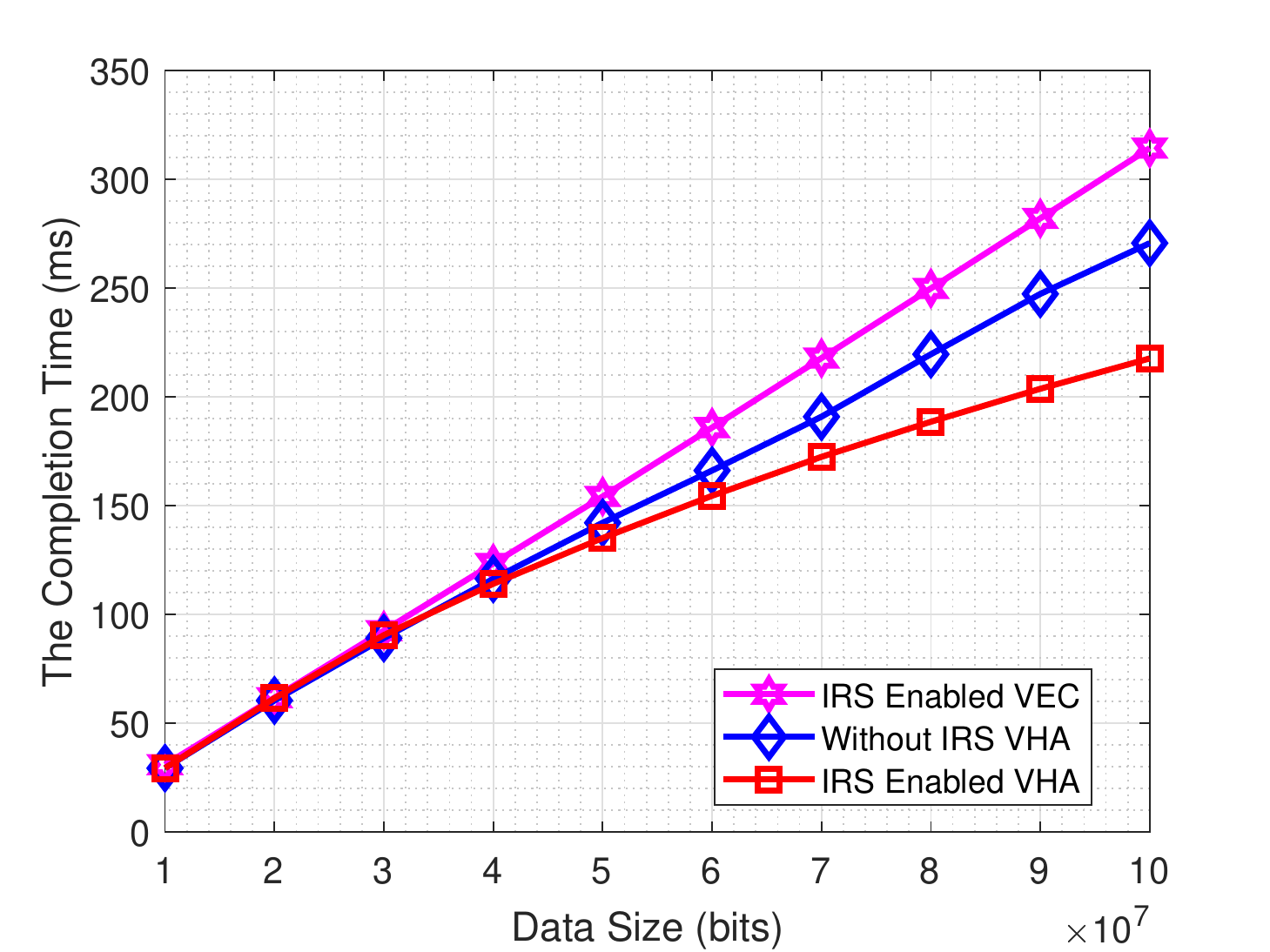}
\caption{Impact of IRS on task's completion time.}
\label{fig:third4}
\end{figure*}

\subsection{System Model, Problem Formulation and Proposed Solution}
We consider a system where $N$ vehicles equipped with low-powered sensors, also called low-powered devices, are randomly distributed over a predefined geographical area. Generally, low-powered devices have minimal computational resources and cannot compute the data requiring extensive computational in a minimal amount of time. As a result, the quality of services of such devices in a high computational vehicular scenario can be highly compromised. To address this issue, vehicles can be efficiently connected to an access point (AP) integrated with a VEC (also stated as VEC AP) through cooperative IRS-enabled drone communications\footnote{In this work, we assume that the direct link between vehicles and VEC AP is nLOS dominant due to large objects. Thus, we consider an IRS enabled cooperative drone communication to assist transmission between vehicles and VEC AP.}, providing communications and computational resources in an on-demand fashion. The drone is flying at a fixed height of $H$, and IRS consists of $K$ passive reflecting elements. The total bandwidth of the system is denoted as $B$. Due to the high
mobility and flexible deployment features, this model considers that IRS is equipped with a drone.
The VEC AP allows vehicles to offload their extensive computation using a partial offloading scheme. The channels between vehicles and VEC AP undergo block fading, i.e., it remains constant for an entire period of time. We also assume that the channel state information is perfectly known in the system. 

The main goal of this optimization framework is to reduce the computational time of the task in the VEC system involving IRS-enabled cooperative drone communication. In particular, this framework optimizes the computational and communication resources among VEC AP and vehicles subject to various practical constraints such as vehicle energy consumption, VEC AP computational resources, drone placement, and efficient phase shift design at IRS, respectively. This framework exploits a partial offloading scheme such that $\Phi_n$ percent of the task is computed locally using local resources while the rest is offloaded to the VEC AP for extensive computation. For instance, the amount of computational resources allocated to vehicle $n$ at the local and VEC AP levels can be represented $\rho_n^l$ and $\rho_n^e$, where $n\in N$. Then the optimization problem of computational task minimization is formulated as non-convex because of the non-linear objective function. Moreover, the coupling of decision variables further complicates the optimization problem. To address the challenges mentioned above and make the optimization more tractable, we perform problem transformation first and then decouple into sub-problems, i.e., drone placement optimization, phase shift control, and computational and communication resources between vehicles and VEC AP. Next, a convex optimization method is used to achieve efficient solution. Specifically, the proposed method is iteratively updated to find the best possible solution that meets all system constraints.
\begin{table}[t]
	\centering
	\caption{Simulation Parameters and their values.}
	\label{tab:simulation_parameters}
	\begin{tabular}{l|l}
\hline 
\hline
Symbol &      Value   \\
\hline
Number of IRS reflecting elements ($K$) &$30$\\
\hline
Number of vehicles ($N$) & $10$ \\
\hline
Height of drone ($H$) & $80$ m \\
\hline
Bandwidth of the system ($B$) & $20$ MHz\\
\hline
Cycles ($c_n$) &$[2\;\;10]$ Kcycles/bits\\
\hline
Data Size ($s_n$) &$[10\;100]$Mbits\\
\hline
Additive white Gaussian noise ($N_o$) & $-173$ dBm\\
\hline
Maximum local computational resources ($\rho_n^{max}$) & $1$ Mcycles/sec\\ 
\hline	 
Maximum edge computational resources ($\rho_e^{max}$) & $25$ Gcycles/sec\\
\hline
\end{tabular}  
\end{table}
\subsection{Numerical Results and Discussion}
To validate the effectiveness of the proposed scheme, extensive simulations are carried out using the parameters given in Table \ref{tab:simulation_parameters}. The numerical results of the proposed IRS-enabled vehicular hybrid approach scheme, i.e., \emph{IRS Enabled VHA} are compared with the following benchmark schemes, i.e., 1) IRS-enabled VEC, i.e., \emph{IRS Enabled VEC}, in which all the tasks of vehicles are offloaded to VEC AP via IRS-enabled drone communication. 2) Without IRS VHA, i.e., \emph{Without IRS VHA} drone communication, in which tasks are partitioned optimally into two portions, one portion of the task is computed locally. At the same time, the other is offloaded to VEC AP without the assistance of IRS-enabled drone communication. 

In the VEC communication network, task segmentation $\Phi_n$ is an important parameter that directly influences the performance of the system. $\Phi_n$ mainly depends on the computational resources of the vehicles, the number of cycle requirements, and the data size of the task as shown in Fig.\ref{fig:figures}.
\subsubsection{Impact of Local computational resources on task segmentation}
Computational resources allocated locally play an essential role while determining the task segmentation variables, as shown in Figure \ref{fig:first}. Results demonstrate that, for the small number of computational resources, the task as a whole is offloaded to VEC AP for extensive computation. This is because local computational resources are not enough to compute it in minimal time. As a result, latency is introduced in a system; hence the quality of service is highly compromised. On the other hand, as the local computational resources increase, the percentage of tasks offloaded decreases, and more tasks are computed locally. 
\subsubsection{Impact of computational cycles requirements on task segmentation}
Similarly, computational cycle requirements are also an important parameter while determining the task segmentation parameter. As perceived from Figure \ref{fig:second}, for the small number of computational cycle requirements, it is efficient to compute the task on the VEC server because of its substantial computational resources. On the other hand, as the computational cycle requirements increase, computational tasks move toward local computation. This trend is because, at the VEC AP, computational cycles are shared among the vehicles, and significant computational cycle requirements demand extensive computation. Whereas shared computation resources at the VEC AP are not enough to meet the high demand, as a result, latency is introduced in a system. To overcome this, local computation is an effective solution to meet the desired requirements, as proved in Figure \ref{fig:second}.  
\subsubsection{Impact of data size on task segmentation} 
\label{DAta}
Following that, Figure \ref{fig:third} demonstrates the impact of data size (bits) on task segmentation. The trend reveals that the task is computed at the VEC AP for the small number of bits. As the data size increases, the computational task shifts toward a local computational scheme. This trend is because offloading a small number of bits consumes less offloading energy than a task's computational energy locally. In contrast, as the data size increases, the offloading energy consumption is more than the local computational energy. So it is efficient to compute the task locally. 
\par
Next, Figure \ref{fig:third4} represents the impact of task segmentation and communication link on the system's performance. The results reveal the proposed scheme \emph{IRS Enabled VHA} outperforms the others by considering computational task time as a performance metric. Results show that for the small number of data sizes, the performance of \emph{IRS Enabled VEC} closely follows the other benchmark schemes. However, as the data size requirements increase, the proposed \emph{IRS Enabled VHA} scheme and \emph{Without IRS VHA} starts performing better. This trend is due to the fact that offloading a large number of bits constitutes more time and energy than the \emph{IRS Enabled VEC} scheme. Whereas in the proposed \emph{IRS Enabled VHA} and \emph{Without IRS VHA} schemes, the task is partitioned into two portions optimally. One portion of the task is computed locally, whereas the other is offloaded to VEC. Likewise, offloading a portion of the task also constitutes less time and energy than complete offloading, compared to \emph{IRS Enabled VEC}.
\par
In addition, the results also demonstrate the IRS's impact on the system's performance. Comparative analysis of \emph{IRS Enabled VEC} and \emph{Without IRS VHA} further reveal the effectiveness of IRS in the VEC communication network. \emph{Without IRS VHA} is a traditional approach in which users offload their portion of the task to VEC AP for extensive computation. At the same time, the presence of an obstacle in the paths results in a weakness in the received signal strength. Therefore, reducing the achievable data rate requires more time to offload the task to VEC AP. Hence, latency is introduced into the system. Whereas in \emph{IRS Enabled VEC} scheme, the drone is equipped with IRS to assist the communication between vehicles and VEC AP, resulting strong signal that helps minimize the offloading time. Overall computational time decreases as a result. 

\section{Open Issues and Future Research Directions}
In this section, we discuss and highlight all open issues and future research directions.
\subsection{Acquisition of Perfect CSI}
Most of the IRS enabled V2X communications works in literature assume that perfect CSI is available at the system while utilizing existing channel estimation techniques. In this regard, the acquisition of highly accurate CSI remains an open issue for V2X communication, especially in the context of IRS. Moreover, three types of channels exist in IRS enabled V2X communications; the first is the direct channel existing between source and destination; in this case, perfect CSI can be obtained with traditional channel estimation techniques executed with ample computational resources at the source side. Furthermore, the second is the indirect channel between source and the IRS, and lastly, the third is the reflection channel constructed between IRS and destination. However, in the second and third cases, obtaining highly accurate CSI is challenging due to the passive nature of IRS elements, which have limited signal processing capabilities. Additionally, the indirect channel estimation is performed by accurately estimating the angle of arrival and departure. In contrast, the reflection channel poses a formidable challenge as environmental conditions, and vehicle location vary over the period of time. Thus, addressing these issues pertaining to the acquisition of accurate CSI is an essential future research direction for IRS enhanced V2X communications.   
\subsection{Machine Learning Techniques}
In future 6G systems, numerous machine learning (ML) techniques are envisioned to perform intrinsic and complicated tasks related to resource allocation and signal processing. In this regard, ML techniques offer a competitive edge over traditional methods or algorithms for seamlessly performing computationally intensive tasks. Moreover, only a handful of works in open technical literature have considered enabling V2X communications via ML techniques despite their wide availability and unlimited capabilities.
Furthermore, research on facilitating IRS enabled V2X communications via ML techniques is a viable futuristic research direction. However, it has not come under the limelight and has not received much attention from the research community. In this context, ML techniques could solve various optimization, classification, prediction, and decision-making tasks which could be extremely useful for facilitating IRS enabled V2X communications.
\subsection{Drone IRS enabled V2X communications}
Due to the mobility feature, Drones can be efficiently used in various civil, public, and military applications. In this regard, both control and communications are essential ingredients of a drone system. Moreover, the upcoming 6G systems will support non-terrestrial networks based on drones serving as aerial BSs or relays. According to 3GPP release 15, a drone flying at an altitude of 80 meters or above has a 100\%  probability of achieving LoS communications capable of mitigating shadowing and signal blockage. Generally, drones are categorized into two types: fixed-wing drones and rotary-wing drones. IRS can be coated on its surface either in a spherical or a flat shape for fixed-wing drones.
On the other hand, for rotary-wing drones, IRS could be mounted on it as a separate module. In this regard, to satisfy quality-of-service requirements for 6G services such as ultra-reliable and low latency communications (URLLC), the specific positioning of IRS enabled drone communication is crucial. Therefore, the deployment of IRS enabled drones to support V2X communication is a promising research direction to investigate.   

\subsection{Optimal beamforming enabling IRS enhanced V2X communications}
In 6G systems, for both non-terrestrial as well as terrestrial networks utilizing IRS, the optimization of beamforming poses a real challenge. In this regard, for non-terrestrial networks based on drone communications, random jittering of drones attributed to strong gusts of wind could lead to an erroneous estimation of the angle of departure existing between IRS and the vehicles. Thus, an accurate method is required to achieve phase alignment of the IRS elements in real-time. On the other hand, for terrestrial networks, the consideration of continuous phase resolution is inaccurate because the IRS phase shifts are discrete in nature. Consequently, such an assumption could lead to signal misalignment, which in turn leads to an inaccurate IRS beamforming. The quantization technique should be applied to continuous phase shifts to address this issue. Thus, as evidenced by the previous discussion beamforming optimization to support IRS enabled V2X communications is a hot futuristic research area.

\subsection{Millimeter wave and Terra Hertz for IRS Enabled V2X Communications}
The growing high data rate demands requires to explore mmWave and THz frequencies ranges to enable vehicular networks to sustain high traffic volumes. The utilization of frequencies in the mmWave and-THz bands is already being investigated and at 120 GHz, data rates of up to 10 Gb/s across distances of up to 850 m have been demonstrated. The road beyond 300 GHz necessitates the creation of innovative transceivers that can function at these incredibly high frequencies. To overcome the intrinsically significant pathloss at these frequencies, such transceivers must have sufficient power and sensitivity while also achieving low noise figures. The penetration losses, high Doppler spread and blocking are some other issues which limits the use of mmWave and THz band in V2X communications. On one hand, the additive advantage of IRS to overcome such issues, makes it a promising candidate to support mmWave and THz bands in V2X communication. On the other hand, the development of RIS to support mmWave and THz bands increases the challenges such as mutual coupling, electromagnetic interference, and many more. Thus, to enable the use of mmWave and THz bands in IRS enabled V2X communications a significant amount research work is required, which makes it a promising future research area.

\subsection{Physical Layer Security in IRS Enabled V2X Communications}
On one hand, the use of RIS enabled wireless communications play a pivotal role in establishing reliable communication links between various entities present in transportation systems. On the other hand, the broadcast nature of wireless signals makes it vulnerable to spoofing assaults, jamming, eavesdropping, etc. Physical layer security plays an active role to overcome such vulnerabilities by improving the average secrecy capacity. Recently, it has been shown that the use of RIS in vehicular network can 
improve the security of legitimate vehicles by improving symbol error probability. Although the literature show some interesting results in terms of utilizing RIS to improve the secrecy capacity, yet the impact of fast optimization of IRS reflection coefficients, advance knowledge of perfect CSI, optimal beamforming, and spectrum allocation are some of open issues that requires attention to find reliable and fast solution in improving physical layer security for IRS enabled V2X communication.


\section{Conclusion}
The IRS has been regarded as an emerging technology in 6G, with the goal of controlling signal propagation and creating a smart radio environment. The IRS is designed to provide LoS-like propagation and energy-efficient communications in 6G V2X networks by leveraging intelligent reflection capabilities. The IRS can help V2X communications in 6G networks by improving multipath propagation and increasing transmission coverage in high spectrum situations. This paper discussed the potential and opportunities of IRS in 6G-empowered V2X communications. In particular, we described different use case scenarios in IRS enabled V2X communications and discussed recent advances. Then, we provided a case study on resource optimization of IRS enabled cooperative drone communication in vehicular edge computing. The numerical results showed the benefits of IRS in terms of task computational time. Finally, we also highlighted current issues and some potential research directions in IRS enabled V2X communications.

\ifCLASSOPTIONcaptionsoff
  \newpage
\fi

\bibliographystyle{IEEEtran}
\bibliography{Wali_EE}

\end{document}